\documentclass[journal]{IEEEtran}
\usepackage{cite}
\usepackage{graphicx,amsmath,amssymb}
\usepackage{subfigure}
\usepackage{citesort}
\usepackage{fancyhdr}
\usepackage{mdwmath}
\usepackage{mdwtab}
\usepackage{balance}
\usepackage{xcolor}
\usepackage{bm}

\usepackage{algorithm}
\usepackage{algorithmic}
\usepackage{multirow}
\usepackage{flafter}
\usepackage{comment}

\usepackage{amsthm}

\newtheorem{remark}{Remark}
\newtheorem{theorem}{Theorem}

\newtheorem{corollary}{\textbf{Corollary}}

\allowdisplaybreaks
\begin{document}

%\title{Fairness of User Clustering in MIMO Non-orthogonal Multiple Access Systems}
%\author{Author 1,  Author 2, Author 3, and Author 4}
%%\markboth{IEEE Transactions on \LaTeX\ }
%%{Hayes}

\title{{STAR-RISs: Simultaneous Transmitting and Reflecting Reconfigurable Intelligent Surfaces}}

%
% {\author{ Yuanwei\ Liu,  Maged\ Elkashlan, Zhiguo\ Ding, and George\ K. Karagiannidis}

 \author{Jiaqi\ Xu, Yuanwei\ Liu~\IEEEmembership{Senior Member,~IEEE}, Xidong\ Mu,~and Octavia A. Dobre~\IEEEmembership{Fellow,~IEEE}.

\thanks{J. Xu and Y. Liu are with the School of Electronic Engineering and Computer Science, Queen Mary University of London, London E1 4NS, UK. (email:\{jiaqi.xu, yuanwei.liu\}@qmul.ac.uk).}
\thanks{X. Mu is with School of Artificial Intelligence and Key Laboratory of Universal Wireless Communications, Ministry of Education, Beijing University of Posts and Telecommunications, Beijing, China. (email: muxidong@bupt.edu.cn).}
\thanks{O. A. Dobre is with Memorial University, Newfoundland and Labrador, Canada. (email: odobre@mun.ca).}
}
\maketitle
%\thispagestyle{fancyplain}
%\pagestyle{fancy}
\begin{abstract}
In this letter, simultaneous transmitting and reflecting reconfigurable intelligent surfaces (STAR-RISs) are studied. Compared with the conventional reflecting-only RISs, the coverage of STAR-RISs is extended to 360 degrees via simultaneous transmission and reflection.
A general hardware model for STAR-RISs is presented. Then, channel models are proposed for the near-field and the far-field scenarios, {base on which the diversity gain of the STAR-RISs is analyzed and compared with that of the conventional RISs.} Numerical simulations are provided to verify analytical results and to demonstrate that full diversity order can be achieved on both sides of the STAR-RIS. 
\end{abstract}

\begin{IEEEkeywords} 
Channel modelling, electromagnetics, performance analysis, reconfigurable intelligent surfaces (RISs), simultaneous transmission and reflection.
\end{IEEEkeywords}

\section{Introduction}
Recently, the concept of smart radio environments and reconfigurable intelligent surfaces (RISs) has received heated discussion in the research community. It is envisioned that the RISs can assist wireless communication networks by intelligently controlling part of the radio environment~\cite{di2019smart,9136592,liu2020reconfigurable}. Specifically, the RIS is a two-dimensional structure that consists of a large number of low-cost elements. These elements are of sub-wavelength sizes and are reconfigurable in terms of their electromagnetic responses. There are diverse applications of introducing RISs into wireless networks, including but not limited to energy efficiency enhancement~\cite{8741198}, multi-user support~\cite{xu2020novel}, visible light communication~\cite{ndjiongue2020use}, and robotic communication~\cite{Mu_robotic}.

{
Despite these advantages, existing research contributions~\cite{liu2020reconfigurable,8741198,xu2020novel,ndjiongue2020use} mainly considered the employment of the reflecting-only RISs, which brings extra topological constraints to the wireless system. Specifically, to receive signals from the reflecting-only RISs, users have to be on the same side as the transmitter. To mitigate this drawback and to facilitate more flexible system designs of the RIS-assisted network, in this letter, we propose the concept of simultaneous transmitting and reflecting RISs (STAR-RISs). From a manufacturing perspective, the proposed new RISs can be implemented using various metasurface-based designs. For example, NTT DOCOMO, INC. recently announced a dynamic metasurface achieving dynamic manipulation of both reflection and penetration in a highly transparent package~\cite{doc}. Other successful implementations achieving manipulation of transmission and reflection are also reported, including graphene-based dynamic metasurfaces~\cite{ZHANG2021374} and multi-layer metasufaces~\cite{wang2018simultaneous}.}

{However, existing research contributions~\cite{doc,ZHANG2021374,wang2018simultaneous} mainly focused on the designing of metasurface prototypes or studying particular physical properties of the designed surface. There is a lack of communication models for the STAR-RISs that are both physically-compliant and mathematically-tractable.} To facilitate the research on STAR-RISs in the field of wireless communications, we separately present a general hardware model and two channel models for the near-field region and the far-field region. Base on the field equivalence principle\cite{liu2020reconfigurable}, the proposed hardware model characterizes each STAR-RIS element by the transmission and reflection coefficients. The channel model gives the closed-form formulation of channel gains in terms of the transmission and reflection coefficients, as well as the geometric settings of the system. {Base on the proposed models, we analyze and compare the diversity gain of the STAR-RISs and conventional RISs. Our numerical results confirm that the STAR-RIS can significantly extend the coverage and achieve full diversity order on both sides of the surface, yielding a superior performance gain over conventional RISs.}

\section{{A General Hardware Model}}\label{sec_hardware}
\begin{figure}[t!]
    \begin{center}
        \includegraphics[scale=0.35]{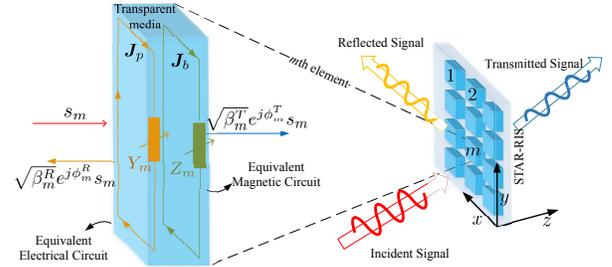}
        \caption{{Schematic illustration of the STAR-RIS.}}
        \label{sys1}
    \end{center}
\end{figure}

{
As shown in Fig.~\ref{sys1}, a schematic representation of the structure of STAR-RISs is presented. Base on the field equivalence principle~\cite{liu2020reconfigurable}, as the STAR-RIS elements are excited by the incident signal, the transmitted and reflected signals can be equivalently treated as waves radiated from the time-varying surface equivalent electric currents $\bm{J}_p$ and magnetic currents $\bm{J}_b$ (also referred to as the bound currents). Within each element, the strengths and distribution of these surface equivalent currents are determined by the incident signal $s_m$ as well as the local surface averaged electric and magnetic impedances $Y_m$ and $Z_m$.
Assume that the STAR-RIS produces both transmitted and reflected signals with the same popularization. At the $m$th element, these signals can be expressed as:
\begin{equation}
s^T_m = T_ms_m,\
s^R_m = R_ms_m,
\end{equation}
where $T_m$ and $R_m$ are the transmission and reflection coefficients of the $m$th element, respectively. According to the law of energy conservation, for passive STAR-RIS elements, the following constraint on the local transmission and reflection coefficients must be satisfied:
\begin{equation}\label{con1}
|T_m|^2+|R_m|^2 \leq 1.
\end{equation}
According to electromagnetic theory, the phase delays of both the transmitted and reflected field are related to $Y_m$ and $Z_m$. In Fig.~\ref{sys1}, the reconfigurability of the element is reflected in the change of the surface impedances, since the transmission and reflection coefficients of the $m$th element is related to the surface impedances as: $
T_m = \frac{2-\eta_0Y_m}{2+\eta_0Y_m}-R_m$, and $R_m = -\frac{2(\eta_0^2Y_m-Z_m)}{(2+\eta_0^2Y_m)(2\eta_0+Z_m)}$, where $\eta_0$ is the impedance of free space~\cite{estakhri2016wave}.
From the perspective of metasurface design, supporting the magnetic currents is the key to achieve independent control of both the transmitted and reflected signals. According to \cite{la2019curvilinear}, single-layered RISs with non-magnetic elements can only produce identical radiation on different sides, which is referred to as the symmetry limitation. 
By introducing the equivalent surface electric and magnetic currents into the model, the proposed hardware model is able to independently characterize the transmission and reflection of each element. To facilitate the design of the STAR-RISs in wireless communication systems, we rewrite these coefficients in the form of their amplitudes and phase shifts as follows:
\begin{equation}
    T_m = \sqrt{\beta^T_{m}}e^{j\phi^T_m},\
    R_m = \sqrt{\beta^R_{m}}e^{j\phi^R_m}, 
\end{equation}
where $\beta^T_{m}, \beta^R_{m} \in [0,1]$ are real-valued coefficients satisfying $\beta^T_{m} + \beta^R_{m} \leq 1$, and $\phi^T_m, \phi^R_m \in [0,2\pi)$ are the phase shifts introduced by element $m$ for the transmitted and reflected signals, $\forall m \in \{ 1,2,\cdots, M \}$.
}

\section{Channel Models} \label{sec_channel}
Base on the proposed hardware model, in this section, we study the communication channels of the STAR-RISs.
As shown in Fig.~\ref{fres}, a downlink STAR-RIS-assisted multi-user wireless network is considered, where the transmitter (Tx) and receivers (Rx) are equipped with a single antenna, and the STAR-RIS consists of $M$ reconfigurable elements. The users are divided into two groups. Users in \textit{group T} are located on the opposite side with the transmitter with respect to the STAR-RIS, and thus, they can only receive signals transmitted by the STAR-RIS. Users in \textit{group R} and the transmitter are located on the same side of STAR-RIS. At the STAR-RIS, signals are transmitted and reflected towards these two groups of users simultaneously. 
Let $h^\chi_k$ denote the direct link between the transmitter and receiver $k$ in \textit{group T} or in \textit{group R}, where the notation $\chi\in\{T,R\}$ is an indicator representing ``$T$'' for receiver in \textit{group T} or ``$R$'' for receiver in \textit{group R}. Let $g^\chi_k$ denote the channel between the transmitter and receiver $k$ through STAR-RIS transmission or reflection.

\subsection{Far-field channel model}
In this case, the receivers are located in the far-field region of the STAR-RIS. We denote the channel between the transmitter and the STAR-RIS by $\bm{h}=(h_1,\cdots,h_M)^T$, where $h_m$ is the channel between the transmitter and the $m$th element. In addition, let $\bm{r}^\chi_k = (r^\chi_{k,1},\cdots,r^\chi_{k,M})^T$ denote the channel between STAR-RIS and receiver $k$ in \textit{group T/R}. Since all receivers are located in the far-field region of the STAR-RIS, the ray tracing technique can be adopted by studying a number of $M$ geometrical rays, each corresponding to a multipath signal propagating through an element. This leads to the channel model as follows:
\begin{equation}\label{gab}
    g^\chi_k = (\bm{r}^\chi_k)^H \text{diag}(\sqrt{\beta_{1}^\chi}e^{j\phi^\chi_m},\cdots,\sqrt{\beta_{M}^\chi}e^{j\phi^\chi_m}) \bm{h},
\end{equation}
where receiver $k$ could be in either \textit{group T} or \textit{group R}. For convenience, we denote $\text{diag}(T_1,T_2,\cdots,T_M)$ by $\bm{\Phi}^T$ and $\text{diag}(R_1,R_2,\cdots,R_M)$ by $\bm{\Phi}^R$. In addition, in the far-field channel model, $\bm{r}^\chi_k$ and $\bm{h}$ can be written in the form of the path loss (large-scale fading) multiplied with the normalized small-scale fading. The large-scale fading depends on the distance between the transmitter, the STAR-RIS, and the receivers, while the small-scale fading depends on the scattering environment. Let $d^\chi_k$ denote the distances between the STAR-RIS and the receivers in \textit{group T/R}, and $d_0$ denote the distance between the transmitter and STAR-RIS. The channel gains can be expressed as:
\begin{equation}\label{sep_1}
    |g^\chi_k| = \frac{1}{(d^\chi_k)^{\alpha_\chi} (d_0)^{\alpha_0}} |({\widetilde r}_k^\chi)^H \bm{\Phi}^\chi\widetilde{\bm{h}}|,
\end{equation}
where ${\widetilde r}_k^\chi$ and $\widetilde{\bm{h}}$ are the corresponding small-scale fading components. In addition, $\alpha_\chi$ and $\alpha_0$ are the path loss coefficients of the corresponding channels.

\subsection{Near-field channel model}
In scenarios where the receivers are located within the near-field region of STAR-RIS, the conventional ray tracing technique based channel models can not be adopted. Base on the Huygens-Fresnel principle~\cite{liu2020reconfigurable}, A. Fresnel and Kirchhoff arrived at the analytical result which is known as the Fresnel-Kirchhoff diffraction formula.
\begin{figure}[t!]
    \begin{center}
        \includegraphics[scale=0.32]{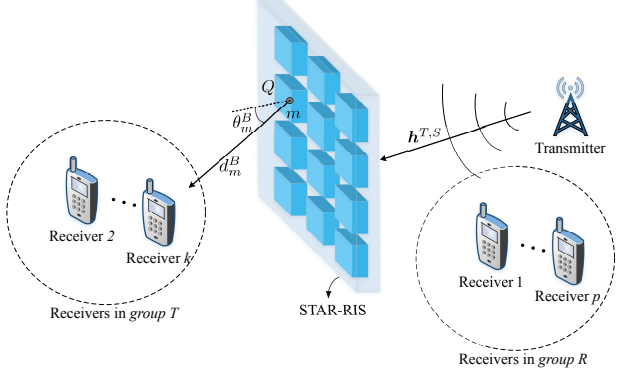}
        \caption{Illustration of the considered RIS-assisted network.}
        \label{fres}
    \end{center}
\end{figure}
As illustrated in Fig.~\ref{fres}, the electromagnetic signal at the receiver can be calculated by summing up the contribution of every elements on the wavefront. The wavefront is chosen as the plane where the STAR-RIS is located.
\begin{equation}\label{int_1}
    g^{\chi}_k = \frac{1}{j\lambda} \iint_{(\Sigma)}U^\chi(Q)F(\theta^{\chi})\frac{e^{2j\pi d^{\chi}_m/\lambda}}{d^{\chi}}d\Sigma,
\end{equation}
where $j$ is the imaginary unit, $(\Sigma)$ denotes a closed surface (wavefront) which contains the RIS elements and the scatters in the environment, $U^\chi(Q)$ is the aperture distributions for the transmitted or reflected waves at point $Q$ on $(\Sigma)$, $F(\theta^{\chi})$ is the leaning factor at point $Q$, $d^{\chi}$ are the distances between $Q$ and the receivers in \textit{group T/R}, and $\lambda$ is the free-space wavelength of the signal.
The integral in \eqref{int_1} can be evaluated element by element. Assuming that the aperture distributions $U^\chi(Q)$ is uniform within each element, at the $m$th element, we have $U^T(Q)=\Phi^T_mh_m$ and $U^R(Q)=\Phi^R_mh_m$. Thus, \eqref{int_1} can be expressed as:
\begin{equation}\label{sum_1}
     g^{\chi}_k = \frac{A_e}{j\lambda} \sum_m \Phi^{\chi}_mh_m F(\theta^{\chi}_m)\frac{e^{2j\pi d^{\chi}_m/\lambda}}{d^{\chi}_m},
\end{equation}
where $A_e$ is the area of each element, and $\theta^{\chi}_m$ is the direction of the users in \textit{group T/R}, with respect to the normal direction of STAR-RIS, as illustrated in Fig.~\ref{fres}. According to the Fresnel-Kirchhoff diffraction formula, the leaning factor is $F(\theta^\chi_m) = (1+\cos\theta^\chi_m)/2$, which holds for both \textit{group T} and \textit{group R}. Thus, the channel gain can be expressed as:
\begin{equation}\label{far_channel_gain}
    |g^{\chi}_k| = \frac{A_e}{\lambda}\Big|\sum_m \Phi^{\chi}_mh_m (1+\cos\theta^{\chi}_m)\frac{e^{2j\pi d^{\chi}_m/\lambda}}{2d^{\chi}_m}  \Big|.
\end{equation}
By comparing these results with the channel gains of the near-field channel model, it can be noticed that the distances between STAR-RIS elements and the receiver can not be treated the same and be brought outside of the summation. In addition, the contribution of the leaning factor should be explicitly considered in the near-field model. 
\vspace{0.1in}
{\begin{remark}Comparing the channel models given in \eqref{sep_1} and \eqref{far_channel_gain}, it can be noticed that the angular distribution of the far-field channel gain is independent of the distance between the receiver and the RIS. However, this is not the case for near-field scenarios. This difference is the essential criteria that set the distinction between the two field regions, and it is further illustrated by our simulation results. According to recent study~\cite{danufane2020path}, the boundary between the near-field and the far-field is $2L_a^2/\lambda$, where $L_a$ is the largest dimension of the RIS aperture.\end{remark}}

\vspace{0.2in}
{
\section{Diversity Analysis}\label{new_sec}
To concisely highlight the benefits of STAR-RISs, in this section, we present a comparison between the following two RISs with specific configurations:\\
\begin{itemize}
    \item \textit{STAR-RIS:} All STAR-RIS elements share a common value for the power ratio between the transmitted signal and the reflected signal. However, the phase delays of each element can be independently configured.
    \item \textit{Conventional RIS:} A composite smart surface consist of reflecting-only and transmitting-only RISs, each with $M_t$ and $M_r$ elements, respectively.
\end{itemize}
}
{
In the following, we compare the performance of the STAR-RIS with the conventional RIS and demonstrate the benefit of deploying STAR-RISs in wireless systems. The power ratio between transmitted and reflected signals is fixed as $\beta^T/\beta^R$, for all elements\footnote{{It is worth mentioning that for STAR-RISs, there is a trade-off between the received power of receivers in \textit{group T} and \textit{group R} by changing the power ratio. Including these power ratios in the optimization can further improve the performance of the system.}}.
Consider a multi-user network assisted by an STAR-RIS or a conventional RIS. We assume that Tx and the STAR-RIS can be positioned with a line-of-sight (LoS) Tx-RIS link. The receivers are all located in the far-field region of the RIS. The direct Tx-Rx links follows Ricean distribution as: $h^{\chi}_k \sim \mathcal{R}(K^{\chi}_d,\Omega^{\chi}_d)$, and the links between each STAR-RIS element and receivers ($r^{\chi}_{k,m}$) are identically and independently distributed, following Ricean distribution as: $r^{\chi}_{k,m} \sim \mathcal{R}(K^{\chi}_s,\Omega^{\chi}_s)$.
According to the proposed far-field channel model for STAR-RISs, the signal-to-noise ratio (SNR) of receiver $k$ for the case of STAR-RIS or conventional RIS can be expressed as $\gamma^{\mu}_k ={|H^{\mu}_k|^2 w^2_k}/{\sigma^2_0}$, where $\mu\in\{S,C\}$ is an indicator representing ``$S$'' for the corresponding quantities in the STAR-RIS case, or ``$C$'' in the conventional RIS case, $H_k$ is the overall channel for receiver $k$, $w_k$ denotes the power allocation for the message of receiver $k$ at the transmitter, and $\sigma_0^2$ is the power spectrum density of the Gaussian white noise at receiver $k$. According to our proposed far-field channel model, the overall channel of receiver $k$ in \textit{group T} or \textit{group R} can be expressed as:
\begin{align}
H^{\mu}_k =&  (\bm{r}^\chi_k)^H \bm{\Phi}^{\chi}_\mu \bm{h}+h^\chi_k,
\end{align}
where $\bm{\Phi}^T_S=\text{diag}(T_1,\cdots,T_M)$,
$\bm{\Phi}^R_S=\text{diag}(R_1,\cdots,R_M)$,  $\bm{\Phi}^T_C=\text{diag}(0,\cdots,0,T_{M-M_t+1},\cdots,T_M)$, and $\bm{\Phi}^R_C=\text{diag}(R_1,\cdots,R_{M_r},0,\cdots,0)$.
For the case of STAR-RIS, under a transmit SNR $\gamma_t$ and target SNR at receiver $\gamma_k$, the outage probability can be formulated as follows:
\begin{equation}
P_{out,S}^{\chi}(k)=Pr\Big\{\Big |\sqrt{\beta^\chi}\sum_{m=1}^M r^{\chi}_{k, m} e^{j\phi_m^{\chi}}+h^{\chi}_k \Big |^2<\frac{\gamma_k\sigma^2_0}{\gamma_0w^2_k} \Big\},
\end{equation}
where $\beta^T/\beta^R$ is the power ratio between transmission and reflection for STAR-RIS elements. For the case of conventional RIS, the outage probability is:
\begin{equation}\label{out_c}
P_{out,C}^{\chi}(k)=Pr\Big\{\Big |\sum_{m=1}^{M_{\chi}} r^{\chi}_{k, m} e^{j\phi_m^{\chi}}+h^{\chi}_k \Big |^2<\frac{\gamma_k\sigma^2_0}{\gamma_0w^2_k} \Big\}.
\end{equation}
Note that the summation in \eqref{out_c} has $M_t$ terms for $k$ in \textit{group T} and $M_r$ terms for $k$ in \textit{group R}. In the following, we compare the best achievable diversity orders of receivers for STAR-RISs and conventional RISs.
\begin{theorem}
For STAR-RISs, the asymptotic outage probability of receiver in \textit{group T/R} can be expressed as follows (superscripts ``$\chi$'' for $K$ and $\Omega$ are omitted):
\begin{align}\label{out1}
\begin{split}
P^{\chi}_{out,S}(k)=&\frac{2^{M+1}(K_s+1)^M(K_d+1)}{(2M+2)! \ \Omega_s^M\Omega_d w_k^{2M+2}}\cdot (\beta^\chi)^{-M/2}\\
&\cdot e^{-MK_s-K_d}\sigma_0^{2M+2}\gamma_k^{M+1}\gamma_t^{-(M+1)}.
\end{split}
\end{align}
For conventional RISs, the asymptotic outage probability of receivers in \textit{group T/R} can be expressed as follows:
\begin{align}\label{out3}
\begin{split}
P^{\chi}_{out,C}(k)=&\frac{2^{M'+1}(K_s+1)^{M'}(K_d+1)}{(2M'+2)!\Omega_s^{M'}\Omega_d w_k^{2M'+2}}\\
&\cdot e^{-M'K_s-K_d}\sigma_0^{2M'+2}\gamma_k^{M'+1}\gamma_t^{-(M'+1)},
\end{split}
\end{align}
where $M'=M_t$ for receivers in \textit{group T} and $M'=M_r$ for receivers in \textit{group R}.
\begin{proof}
Following the assumption of Ricean fading channels, the probability density functions (PDFs) of Tx-Rx direct link and RIS-Rx links can be expressed in the following series form:
\begin{align}
p_{|r^{\chi}_{k,m}|}(x)=\frac{2(K^{\chi}_s+1)}{\sqrt{\beta^\chi}\Omega^{\chi}_s}e^{-K^{\chi}_s}x+o(x^2),\\
p_{|h_k^{\chi}|}(x)=\frac{2(K^{\chi}_d+1)}{\Omega^{\chi}_d}e^{-K^{\chi}_d}x+o(x^2),
\end{align}
where $o(\cdot)$ is the little-O notation, and $o(x^2)$ denotes a function which is asymptotically smaller than $x^2$.
To obtain the PDF of the magnitude of the overall channel ($|H_k^{\mu}|$), we calculate its Laplace transform according to the convolution theorem. Consider the case of STAR-RIS, we have:
\begin{align}
\begin{split}
\mathcal{L}\{ p_{|H_k^{E}|}(x) \}(t)=\big(\mathcal{L}\{ p_{|r^{\chi}_{k,m}|(x)} \} \big)^M\mathcal{L}\{ p_{|h_k^{\chi}|}(x) \}\\
=\frac{2^{M+1}(K_s+1)^M(K_d+1)}{(\beta^\chi)^{M/2}\Omega_s^{M}\Omega_d} e^{-MK_s-K_d}\cdot t^{-2M-2}.
\end{split}
\end{align}
Thus, by applying the inverse Laplace transform, the outage probability can be calculated according to:
\begin{equation}
P^{\chi}_{out,S}(k) = \int_{0}^{\frac{\sigma_0}{w_k}\sqrt{\frac{\gamma_k}{\gamma_t}}}p_{|H_k^{E}|}(x)dx,
\end{equation}
The proof for the case of conventional RISs follows the same strategy, thus is omitted here for brevity.
\end{proof}
\end{theorem}
In the following remark, we consider the diversity order defined through outage probability as follows:
\begin{equation}
d^{\chi}(k) = -\lim_{\gamma_t \to \infty}\frac{\log P^{\chi}_{out}(\gamma_k)}{\log \gamma_t}.
\end{equation}
\begin{remark}\label{diversity_order}
According to \textbf{Theorem 1}, the diversity orders of different receivers can be readily obtained from \eqref{out1} and \eqref{out3}. For STAR-RIS, the achievable diversity orders of receivers in both \textit{group R} and \textit{group T} are $d^T_{S} = d^R_S=M+1$. Thus, the combined diversity order is $d^T_{S} + d^R_S=2M+2$. For conventional RISs, the achievable diversity order for receivers in \textit{group T} and \textit{group R} follows: $d^T_{C} + d^R_C=M+2$, yielding smaller achievable diversity orders compared to STAR-RISs.
\end{remark}
}
\begin{figure*}[t!]
\centering
\subfigure[Simulation setup]{\label{n0}
\includegraphics[width= 2.3in]{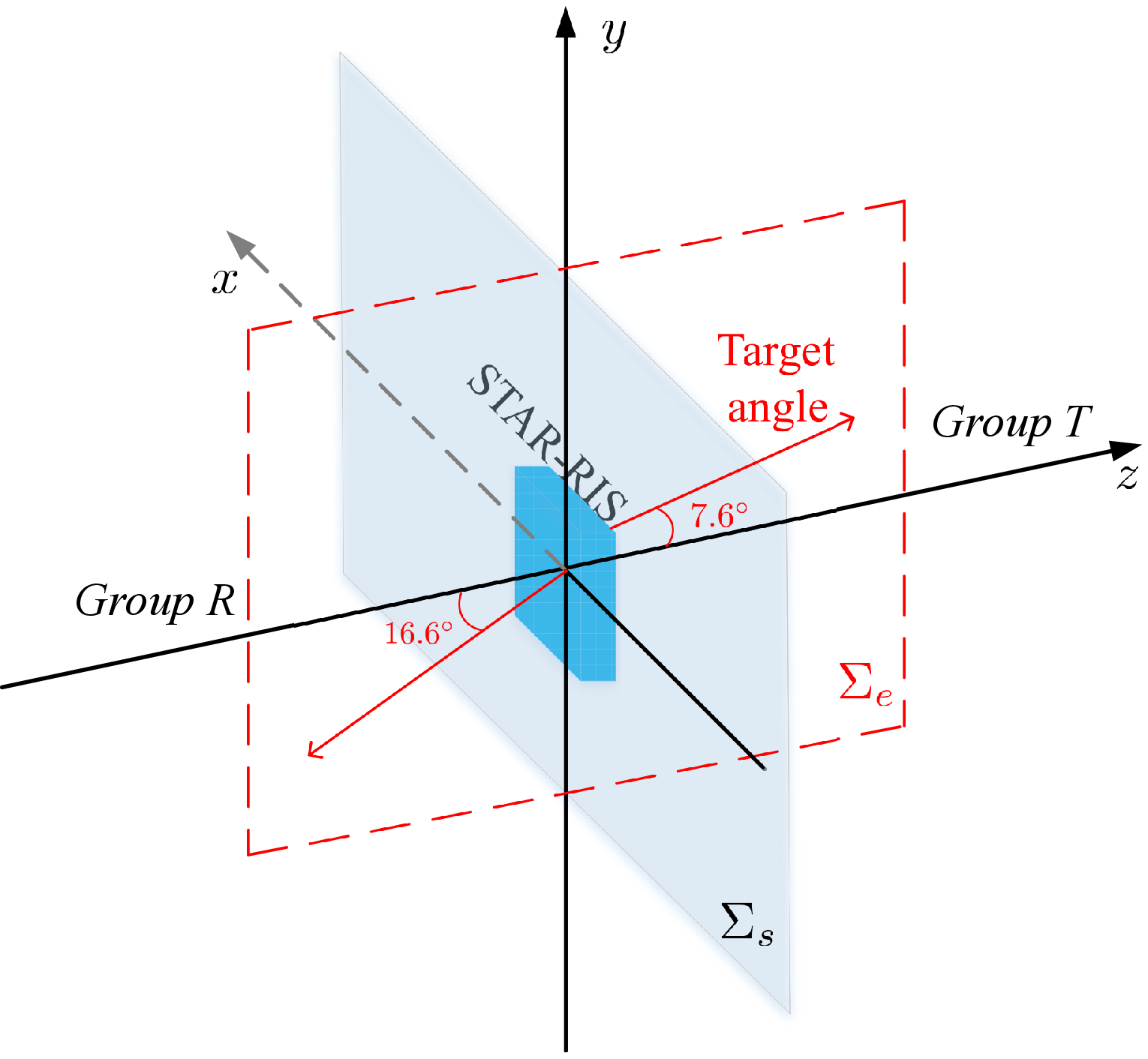}}
\subfigure[STAR-RIS]{\label{na}
\includegraphics[width= 2.3in]{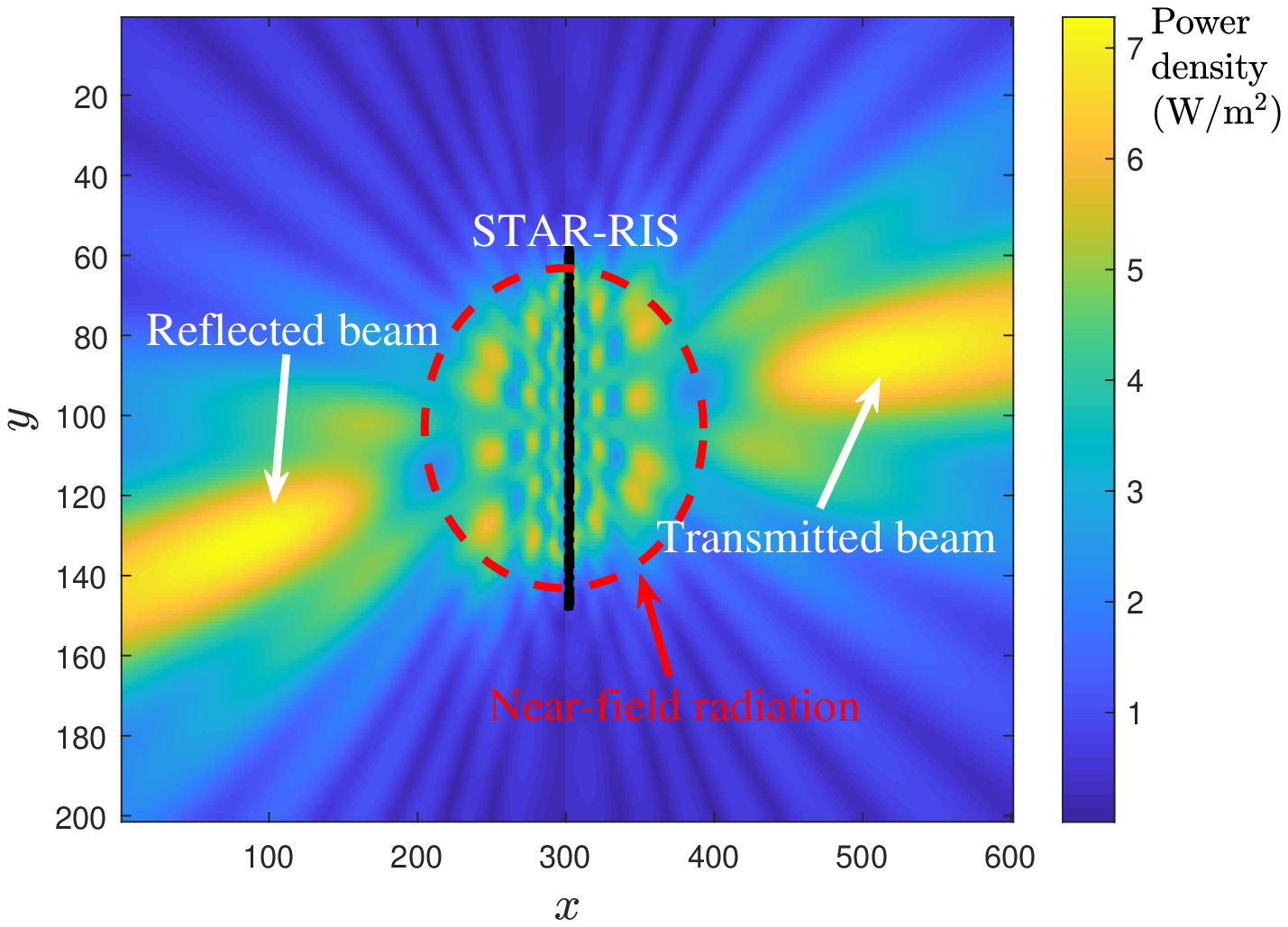}}
\subfigure[Conventional RIS]{\label{nb}
\includegraphics[width= 2.3in]{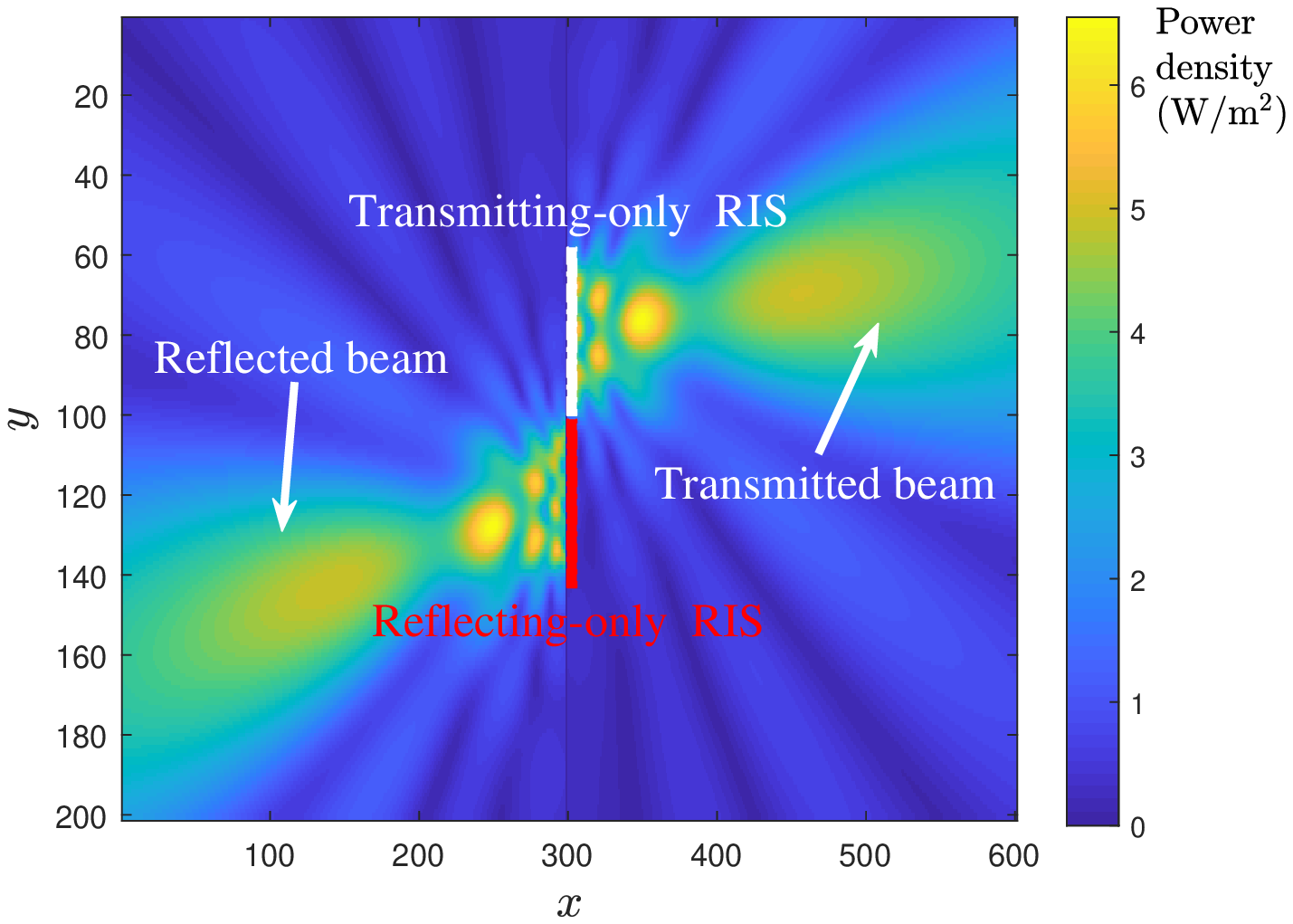}}
\caption{{Illustration of the simulation setup and simulated radiation coverage observed on plane $\Sigma_e$.}}\label{nice}
\end{figure*}
\section{Numerical Results}\label{num}
\begin{figure}[t!]
    \begin{center}
        \includegraphics[scale=0.52]{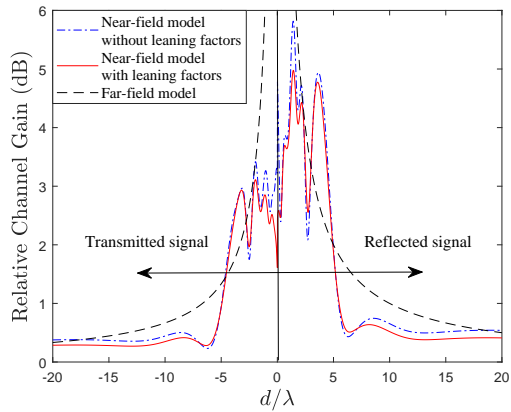}
        \caption{Channel gain computed using far-field and near-field channel models.}
        \label{sca}
    \end{center}
\end{figure}

{
In this section, we numerically demonstrate our proposed models to compare the coverage of the STAR-RISs and the conventional RISs. Then, we simulate the outage probabilities of receivers severed by STAR-RISs and the conventional RISs to verify our analytical results.
The geographical setup is shown in Fig.~\ref{n0}, where the RIS is positioned at the origin in the $z=0$ plane. We simulate a $16\times16$ STAR-RIS, where spacing of each element is chosen as $\lambda/2$, and a conventional RIS with $M_t=M_r=128$.}\\
\indent {In Fig.~\ref{nice}, we compare the coverage of the proposed STAR-RIS with conventional RISs by simulating their radiation patterns. The simulation is carried out as follows: Base on~\eqref{int_1}, we calculate the channel gain of each point on the plane $\Sigma_e$. For STAR-RIS, we set $\beta^T =\beta^R=1$. The phase shift configurations of both the STAR-RIS and conventional RIS is chosen according to the cophase condition~\cite{liu2020reconfigurable}. Specifically, the target angles of the transmitted signal and the reflected signal are set as $7.6^\circ$ and $16.6^\circ$, respectively. As illustrated in Fig.~\ref{na}, the power density of the STAR-RIS is significantly higher near the target angles, forming a beam-like radiation pattern. In contrast, in Fig.~\ref{nb}, the transmitted and reflected beams of the conventional RIS are less focused with weaker channel gains. This observation is consistent with our analytical results presented in Section.~\ref{new_sec}. Moreover, it can be observed that the angular distribution of power density has an irregular pattern which highly depends on the distance in the near-field region, while in the far-field region, a fixed radiation pattern can be observed. These observations of the radiation patterns are consistent with the results of the two proposed channel models.}\\
\indent Fig.~\ref{sca} shows the channel gain obtained by the $16\times16$ STAR-RIS. The power ratio between transmission and reflection is set to $2$:$3$. For the horizontal axis, $d$ denotes the distance between the receiver and the center of STAR-RIS. The signal is reflected to the region where $d>0$, and transmitted to the region where $d<0$. In the simulation, we assume that the receiver moves on both sides of STAR-RIS, along a straight line in $\Sigma_e$ with an angle of $60^\circ$ with respect to the normal direction of STAR-RIS. As illustrated in Fig.~\ref{sca}, the dash-dotted blue line represents the channel gains calculated using the near-field channel model without considering the leaning factor. In other words, we assume $F(\theta^T_m)=F(\theta^R_m)=1$. The solid red line represents the channel gains calculated using the near-field channel model with leaning factor $F(\theta^\chi_m) = (1+\cos\theta^\chi_m)/2$. The dashed black line represents the channel gains calculated using the far-field channel, according to \eqref{sep_1}. It can be observed that results for the near-field model start to approach the far-field model after the receiver is several wavelength away from STAR-RIS. Within the near-field region, the far-field model fails to provide a physically meaningful result as the channel gain tends to infinity when the distance approaches zero.
\begin{figure}[t!]
    \begin{center}
        \includegraphics[scale=0.52]{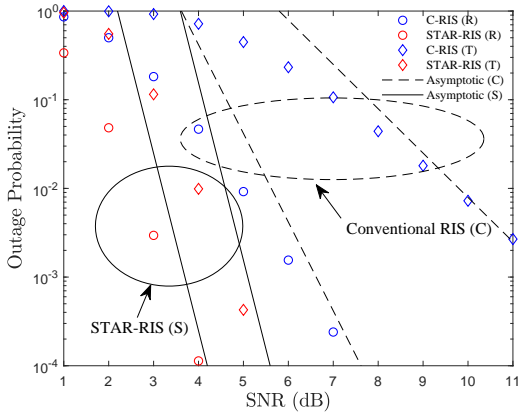}
        \caption{{Outage probabilities of receivers served by STAR-RIS and conventional RIS (C-RIS).}}
        \label{outage}
    \end{center}
\end{figure}
{
\indent Fig.~\ref{outage} illustrates the outage probabilities and diversity orders of STAR-RIS and conventional RIS. The red markers represent the simulated outage probability for receiver assisted by an STAR-RIS with eight elements ($M=8$) and a power ratio of $2$:$3$. The blue markers represent a conventional RIS with $M_t=3$ and $M_r=5$. It can be observed that the simulated outage probabilities fit well with the analytical asymptotic results. The diversity orders of receivers in \textit{group T/R} are consistent with our analytical results. For STAR-RIS, receivers in both groups achieves full diversity order of $M+1$, for conventional RIS, the diversity orders follow $d^T_{C} + d^R_C=M+2$, which is consistent with \textbf{Remark \ref{diversity_order}}.
}
\section{Conclusions}
{
In this letter, the concept of STAR-RISs is proposed.
A general hardware model and two channel models were proposed. To evaluate the performance of STAR-RISs, we derived expressions for the asymptotic behaviour of the outage probability. Numerical simulations demonstrated the differences between the channel gains in the far-field and the near-field region and verified our proposed channel models. It is also demonstrated that the proposed STAR-RISs can extend coverage and achieve higher diversity order on both sides of the surface, compared to conventional RISs. Base on the proposed models, the joint optimization of transmission and reflection beamforming of the STAR-RISs constitutes an interesting topic for future research.
}

\bibliographystyle{IEEEtran}
\bibliography{mybib}

\end{document}